 \documentclass[%
   ,twocolumn%
  , aps
  , prl
 , showpacs
 ]{revtex4}

\usepackage{amsmath}
\usepackage{amssymb}
\usepackage{bm}
\usepackage{times}

\newcommand{\textfrac}[2]{\ensuremath{#1/#2}}
\newcommand{\braket}[1]{\ensuremath{\left\langle#1\right\rangle}}
\newcommand{\rmd}{{\mathrm{d}}}
\newcommand{\Int}[3]{\int\limits_{#1}^{#2}\!\rmd#3\;}

\newcommand{\Figref}[1]{Figure~\ref{#1}}

\newcommand{\Eqref}[1]{Equation~\eqref{#1}}

\newcommand{\Refcite}[1]{Ref.~\citep{#1}}

\usepackage{graphicx}
\newcommand{\gpfig}[1]{%
  \includegraphics[width=8.6cm]{#1} 
}
\usepackage{psfrag}
\newcommand{\xlabel}[1]{\psfrag{xlabel}[t][B]{#1}}
\newcommand{\ylabel}[1]{\psfrag{ylabel}[b][t]{#1}}

\newcommand{\Ninf}{\ensuremath{N\to\infty}}

\newcommand{\Rg}{\ensuremath{R_{\mathrm{g}}}}
\newcommand{\Rh}{\ensuremath{R_{\mathrm{H}}}}
\newcommand{\Ri}{\ensuremath{R_{\mathrm{I}}}}

\newcommand{\Rgth}{\ensuremath{\Rg^{\theta}}}

\newcommand{\hsH}{\ensuremath{h^{*}}}
\newcommand{\zs}{\ensuremath{z^{*}}}
\newcommand{\ds}{\ensuremath{d^{*}}}

\newcommand{\alfag}{\ensuremath{\alpha_{\mathrm{g}}}}
\newcommand{\alfaI}{\ensuremath{\alpha_{\mathrm{I}}}}
\newcommand{\alfaH}{\ensuremath{\alpha_{\mathrm{H}}}}

\newcommand{\Urd}{\ensuremath{U_{\mathrm{R D}}}}
\newcommand{\Urdth}{\ensuremath{U_{\mathrm{R D}}^{\theta}}}

\newcommand{\Dk}{\ensuremath{D_{\mathrm{K}}}}
\newcommand{\nueff}{\ensuremath{\nu^{\mathrm{eff}}}}

\newcommand{\urlprefix}{\relax}
\begin{document}
\title{The origin of dynamic scaling in dilute polymer solutions}

\author{P. Sunthar}
\email{P.Sunthar@gmail.com}
\author{J. Ravi Prakash}
\email{Ravi.Jagadeeshan@eng.monash.edu.au}
\affiliation{Department of Chemical Engineering, Monash University,
  Melbourne, VIC 3800, Australia}
\date{\today}

\begin{abstract}
  The hydrodynamic radius of a polymer chain, obtained using Brownian
  dynamics simulations of the continuum Edwards model, is found to
  obey a crossover in the excluded volume parameter $z$, which is
  significantly different from that observed for the radius of
  gyration.  It is shown that this difference arises from
  contributions due to dynamic correlations to the diffusivity, which are
  ignored in the commonly used definition of hydrodynamic radius based
  on the Kirkwood expression. The swelling of the
  hydrodynamic radius from the $\theta$-state, obtained from
  simulations, shows remarkable agreement with experimental
  measurements.
\end{abstract}

\maketitle


The mean size of a polymer in a dilute solution can be obtained by
static measurements that yield the radius of gyration \Rg\ or by
dynamic experiments that provide the diffusion coefficient $D$.  An
outstanding problem yet unresolved is the observed difference in the
scaling with molecular weight between static and dynamic measurements
\citep{dgen79}.  For instance, in the limit of large molecular weight
$M$, \Rg\ scales as $M^{0.59}$, while the hydrodynamic radius $\Rh
(\propto D^{-1})$ scales as $M^{0.57}$ \citep{graetal99}.  In
addition, in the crossover regime between the $\theta$-state and the
good solvent limit, the growth of \Rg\ and \Rh\ relative to the
$\theta$-state, in terms of the scaling variable $z = v_{0} \, ( 1 -
T_{\theta} / T ) \, \sqrt{M}$, is observed to be significantly
different.  Here, $v_{0}$ is a chemistry dependent prefactor, $T$ is
the temperature, and $T_{\theta}$ is the temperature at the
$\theta$-state.  Theoretical attempts to explain the origin of this
anamalous dynamic scaling behavior have mainly based the definition
of \Rh\ on the Kirkwood expression for diffusivity
\citep{WeiClo79,BenAkc79,DouFre84a,DouFre84b}. In this work, we show
using exact Brownian Dynamics simulations of the continuum Edwards
model \citep{Edw64}, that (i) the use of Kirkwood's expression in fact
leads to nearly identical scaling in the two cases, (ii) in order to
exhibit dynamic scaling that is distinct from static scaling, \Rh\ 
should be based on the long-time diffusivity, and (iii) the latter
definition of \Rh\ leads to excellent agreement with experimental
observations.

Theoretical treatments of the static and dynamic properties of
polymers in a dilute solution are typically carried out with a
bead-spring chain model, which consists of $N$ beads connected by
linear springs.  The swelling of a polymer in a good solvent is
usually represented in terms of the ratio of the size in the presence
of excluded volume (EV) interactions between the beads to size in the
$\theta$-state.  The swelling of the radius of gyration is denoted as,
$\alfag = \Rg/\Rgth$ and that of the hydrodynamic radius as $\alfaH =
\Rh/\Rh^{\theta}$.  The hydrodynamic radius \Rh\ is often approximated
by a static measure constructed from the Kirkwood estimate of the
diffusivity \cite{Kir54}
\begin{equation}
  \label{eq:dk}
  \frac{1}{\Rh} \sim \Dk = \frac{1}{4} \left( \frac{1}{N} +
  \sqrt{\pi} \, \hsH \, \frac{1}{\Ri} \right),
\end{equation}
where, $\Ri$ is the 
inverse radius
(or sometimes known as the ``dynamic'' radius):
\begin{equation}
  \frac{1}{\Ri} \equiv \frac{1}{N^{2}} 
  \, \sum_{\substack{\mu,\nu=1 \\ \mu \neq \nu}}^{N} \,
  \braket{\frac{1}{r_{\mu\nu}} }. 
\end{equation}
Here, $r_{\mu\nu}$ is the distance between the beads $\mu$ and $\nu$.
It is observed from experiments that $\alfaH < \alfag$, and for the
typical maximum molecular weights studied, the effective scaling
exponent (defined as $\nueff = 1/2 + \partial \log\alpha/\partial \log
M$) is $\nueff = 0.59$ for \alfag\ and $\nueff = 0.57$ for \alfaH.  An
early explanation of this anomalous dynamic scaling behavior employed
a model of self-excluding Gaussian blobs \citep{WeiClo79} to show that
the segments within the Gaussian blob have more influence on $1/\Ri$
than on $\Rg$.  It was shown that this leads to a swelling in \Ri,
defined as $\alfaI = \Ri/\Ri^{\theta}$, which obeys $\alfaI < \alfag$.
The effective exponent for \Ri\ was shown to reach the asymptotic
value of $\nu = 0.59$ much more slowly than that for \Rg, and an
approximate estimate provided $\nueff = 0.57$ for \Ri\ for the
experimental system considered. This idea and other modified blob
theories have subsequently been discounted by Monte Carlo simulations
\citep{SchBau86,LadFre92} of excluded volume chains which show that
the model ignores the excess swelling of internal parts of the chain
and that in fact the effective exponent is significantly closer to
0.59 than 0.57 \citep{Schafer99}, a fact supported by renormalisation
group (RG) calculations \citep{DouFre84a} and our own results shown
here.

The crossover scaling behaviour (from the $\theta$-limit to the
good-solvent limit) of static properties, such as \alfag\, is a well
studied problem.  One of the most successful and widely used
approaches is the Edwards continuous chain (\Ninf) representation of
the bead-spring model \citep{Edw64}.  RG calculations have shown that
this model correctly captures the static scaling behavior for
temperatures asymptotically near the $\theta$-temperature, in close
agreement with experimental observations
\cite{DouFre84a,Schafer99,CCJ85,MutNic87}.  However, RG calculations
\citep{DouFre85} of the crossover of dynamic swelling ratios based on
the definition of the Kirkwood diffusivity lead to $\alfaI \approx
\alfag$ (to within 1\%).

An alternative explanation for the lower value of \alfaI\ is the
existence of a draining effect in hydrodynamic interactions
\citep{DouFre84b}. This implies the introduction of another
(draining) parameter into the theory.  However, a recent compilation
of dynamic data \citep{Tometal02} shows a significant degree of
universality in the crossover in $z$, clearly indicating the
absence of another parameter (or that the chains are non-free
draining).
  
In this work, we have reexamined the dynamic crossover problem,
(restricting our attention to the Edwards' continuum chain
description) in the light of three recent developments.  First is the
introduction of a scheme to simulate the continuum Edwards model using
the Brownian dynamics (BD) method \citep{SatPra03}, which permits the
study of dynamical and rheological properties.  In this scheme, a
repulsive Gaussian EV potential is used to mimic the $\delta$-function
potential in the Edwards model, and the mapping of the model
parameters to the EV parameter $z$ is shown to occur in the continuous
chain limit \cite{Pra01b}.  The second is the advance made in the
treatment of fluctuating HI using fast approximation methods
\citep{jenetal00}.  The third is the resolution of a long standing
problem with regard to the diffusivity of a polymer chain.
\citet{LiuDun03} have numerically proven that for chains with
fluctuating HI, the long time diffusivity is lower than the short time
Kirkwood estimate \Dk, and is given by $D = \Dk - D_{1}$, where
$D_{1}$ is the equilibrium time correlation function of the drift term
in the stochastic differential equation governing the motion of the
beads of the chain \citep{Fix81}.  These developments permit us to
compute the diffusivity from the Edwards model at a finite $z$ and
obtain the crossover exactly in the non-free draining limit.  We show
that in the crossover regime $\alfaI \approx \alfag$ , and \alfaH\ 
obtained from the long-time diffusivity is able to reproduce the
observed anomaly in the swelling of hydrodynamic radius.


To simulate the continuum Edwards model, we begin with a discrete
bead-spring chain with $N$ beads subject to Brownian motion.  The
dimensionless displacement of a bead position $\Delta R_{\mu i}$
during a finite time increment $\Delta t$ is taken to be
\citep{SunRav05}:
\begin{equation}
  \label{eq:sde}
  \Delta R_{\mu i} = 
  \frac{1}{4} \, D_{\mu \nu i j} \, F_{\nu j}\, \Delta t + 
  \frac{1}{\sqrt{2}} \, B_{\mu \nu i j}\Delta W_{\nu j} ,
\end{equation}
where, the summation convention is implied for repeated indices.  and
a Greek subscript $\mu,\nu,\ldots = [1,2,\ldots,N ]$ denotes the bead
index and a Roman subscript $i,j,\ldots = [1,2,3]$ denotes a Cartesian
coordinate.  Here, $D_{\mu\nu i j}$ is the diffusion tensor, $F_{\mu
  i}$ is a matrix of the total body force acting on the beads, $B_{\mu
  \nu i j }$ is taken as the square root matrix of $D_{\mu\nu i j}$
defined as $B_{\mu \theta i k} B_{\nu \theta j k} \equiv D_{\mu \nu i
  j}$, and $\Delta W_{\mu i}$ is an increment to the Weiner process
$W_{\nu j}$.  In the absence of EV interactions, the distribution of
bead connector vectors is assumed to be Gaussian.  A repulsive
narrow-Gaussian potential is used to regularise the $\delta$-function
EV potential \citep{Pra01b} considered in the Edwards model
\citep{Edw64}. The potential is parameterised by \zs\ representing the
strength and \ds\ representing its width \citep{SatPra03}.  The
solvent quality (EV parameter) is given by $z=\zs \, \sqrt{N}$,
obtained in the simultaneous limit of \Ninf\ and $\zs \to 0$.  In
\Refcite{SatPra03} the expression $\ds = K \, {\zs}^{1/5}$ (where $K$
is a constant) was found to be a computationally efficient way to
reach the limit $\ds \to 0$, which recovers the $\delta$-function.
This method was shown to accurately predict the swelling of the
gyration radius\citep{SatPra03} observed in experiments
\citep{MiyFuj81}.  To simulate hydrodynamic interactions (HI) between
the beads we have used the usual Rotne-Prager-Yamakawa expression for
the HI tensor, and an accelarated method of finding the square root
matrix $B_{\mu \nu i j}$ \citep{jenetal00}.  These and other details
of the simulation are discussed elsewhere \citep{SunRav05}.

\citet{Fix81} established from a preaveraged treatment for Gaussian
chains that dynamic correlations play a minor role in the diffusivity
of chains.  He showed that the long time diffusivity is given by $D =
\Dk - D_{1}$, where the contribution $D_{1} > 0$ is related to the
time-correlation function of the drift term in \Eqref{eq:sde}, and is
only about 2\% of \Dk.  Defining $A_{i} = (\sum_{\mu} D_{\mu \nu i j}
\, F_{\nu j})/4 $, we have
\begin{equation}
  D_{1} = \frac{1}{3\,N^{2}} \, \Int{0}{\infty}{t} \braket{A_{i}(0)
  \, A_{i}(t)}  .
\end{equation}
\citet{LiuDun03}, who recently performed simulations to obtain
diffusivity $D$ directly, and through the Fixman's relationship,
effectively proved the relation by showing numerically (for excluded
volume chains) the variation of diffusivity from its short time value
\Dk\ to the long time limit.  Even in this case, it was found that the
contribution of $D_{1}$ is only a small fraction of $D$.  This is one
of the prime reasons for the neglect of dynamic correlations in the
diffusivity and the use instead, of the Kirkwood estimate \Dk---an
approximation which we show below has gone unnoticed in attempts
to explain dynamic scaling.

We have validated Fixman's expression in $\theta$-conditions with
fluctuating HI, by carrying out similar simulations to that in
\Refcite{LiuDun03}. In this work, we have used the computationally
easier method of evaluating $D$, which is to evaluate the static
average \Dk\ and obtain $D_{1}$ from the time-correlation function.
This immediately enables one to study the crossover of both the static
estimate and the dynamically correlated value.

We define two swelling ratios for the two different size measures
obtained from the diffusivity.  The commonly used Kirkwood estimate
gives the swelling of the inverse radius $ \alfaI =
\textfrac{\Ri}{\Ri^{\theta}} = \textfrac{\Dk^{\theta}}{\Dk}$, in which
the second equation is obtained in the limit \Ninf\ (which is true for
the continuum model).  The hydrodynamic radius obtained from the
long-time diffusivity has a swelling ratio $\alfaH =
\textfrac{D^{\theta}}{D}$.  \alfaI\ is obtained accurately from
free-draining runs by carrying out simulations at two different values
of $K=1.0$ and 1.5.  \alfaH\ is obtained from separate runs
incorporating fluctuating HI and is carried out for two values of HI
parameter (which is the dimensionless radius of a bead), $\hsH =
0.19$ and 0.30.  The results from simulations performed at finite $N$
are extrapolated to \Ninf, keeping $z$ and \hsH\ constant.  This limit
corresponds to the continuum Edwards model in the non-free draining
limit ($\hsH\,\sqrt{N} \to \infty$).


\begin{figure}[tbph]
  \centering

  \xlabel{$1/\sqrt{N}$}
  {\ylabel{$\alfaI$} 
    \psfrag{K = 1.0}{$K=1.0$}
    \psfrag{K = 1.5}{$K=1.5$}
    \gpfig{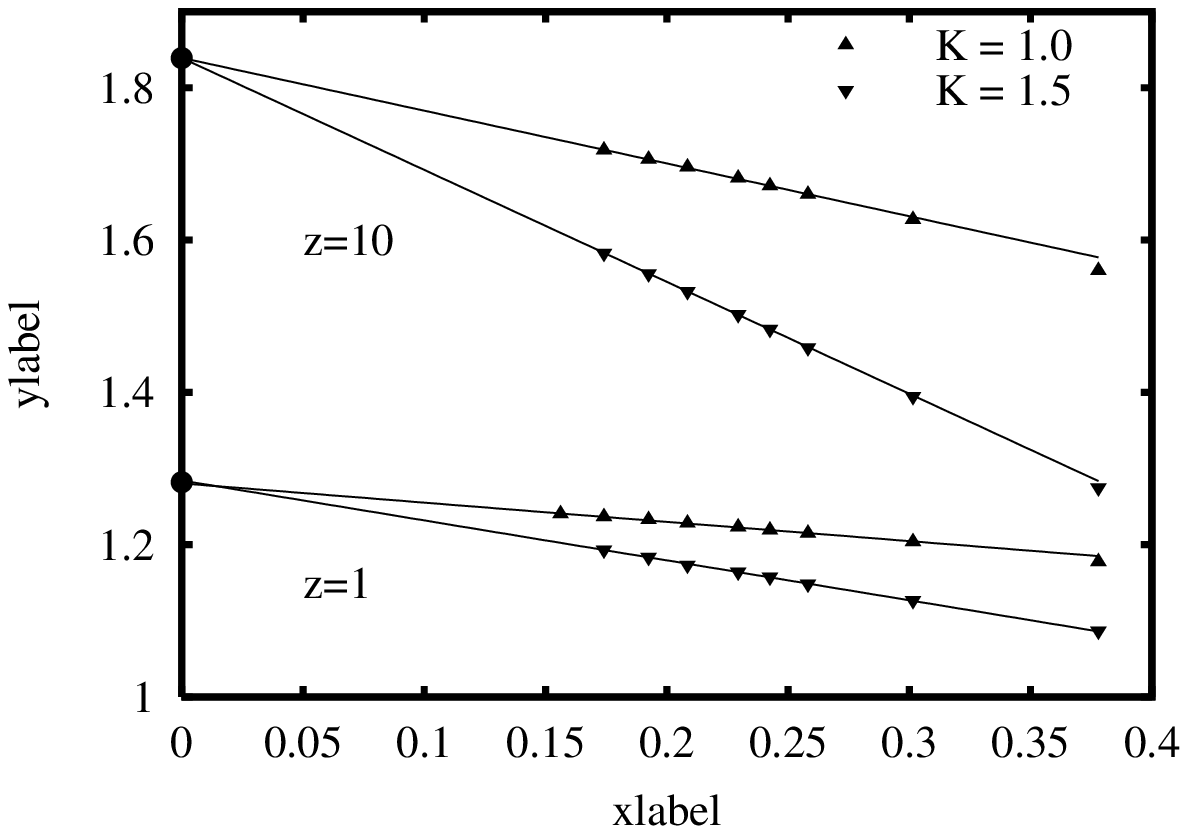} }
  {\ylabel{\Urd} 
    \psfrag{0.19}{$\hsH = 0.19$}
    \psfrag{0.30}{$\hsH = 0.30$}
    \gpfig{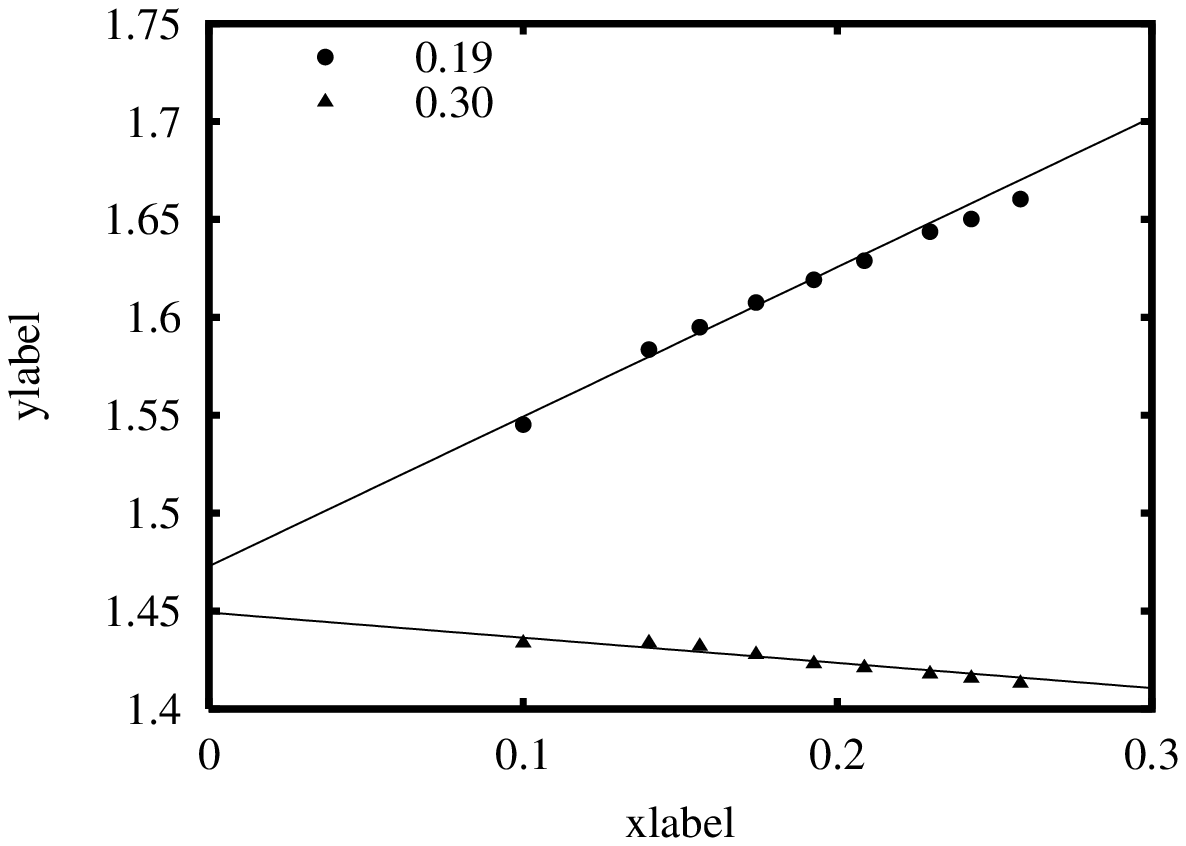} }
  \caption{Universality at a fixed solvent quality $z$
    in the continuous chain limit \Ninf. (a) The independence of the
    extrapolated value of \alfaI\ from the width of the Gaussian
    repulsive potential and (b) A similar independence of \Urd\ from
    the HI parameter \hsH at a solvent quality $z=1$.}
  \label{fig:aIHz-ext}
\end{figure}

\begin{figure}[tbph]
  \centering
  {
    \xlabel{$z$}
    \ylabel{$\alfag; \alfaI; \alfaH$}
    \psfrag{ag}{\alfag}
    \psfrag{aI}{\alfaI}
    \psfrag{aH}{\alfaH}
    \gpfig{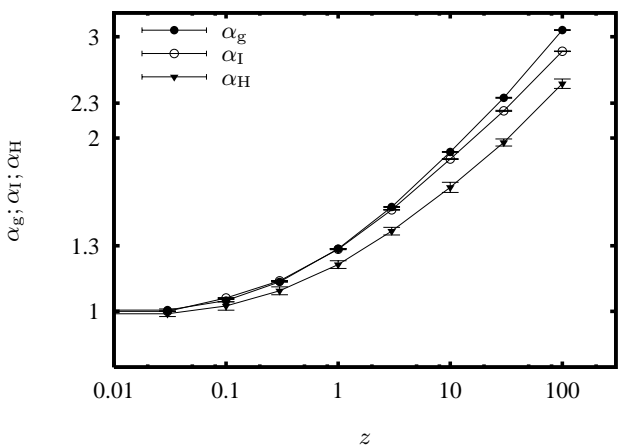}
  }

  \caption{Universal solvent quality crossover of swelling of two
    measures of polymer size based on diffusivity (inverse radius
    \alfaI\ and hydrodynamic radius \alfaH) compared with the
    crossover of swelling of the radius of gyration (\alfag).}
  \label{fig:aIHz}
\end{figure}

The universal nature of the swelling ratios \alfaI\ and \alfaH\ is
seen in \Figref{fig:aIHz-ext}.  The independence of the extrapolated
value of \alfaI\ from the width of the Gaussian repulsive potential
indicates $z$ is the only relavant parameter. A similar behaviour is
observed for \alfaH\ (not shown here).  To calculate \alfaH\ we find
it convenient to use the following expression
\begin{equation}
  \alfaH = \frac{\Rh}{\Rg} \frac{\Rg}{\Rgth}
  \frac{\Rgth}{\Rh^{\theta}} = \Urd^{-1} \, \alfag \, \Urd^{\theta} ,
\end{equation}
where, $\Urd \equiv \Rg/\Rh$. This separates out the weak dependence
of \alfaH\ on \hsH\ at finite $N$ into the two strong dependences of
each of \Urd\ and $\Urd^{\theta}$. \alfag\ is independent of \hsH\ and
can be obtained accurately from free-draining simulations along with
\alfaI. Our
simulations of Gaussian chains with fluctuating HI yield a value of
$\Urdth = 1.38 \pm 0.01$.  \alfag\ is determined from free-draining
simulations \citep{SatPra03}.  \Urd\ is determined by carrying out
simulations at constant \hsH, and the extrapolations to \Ninf\ are
carried out for two values of $\hsH = 0.19$ and $0.30$ as shown in
\Figref{fig:aIHz-ext}b.  The two extrapolations can be seen to
approach a common limiting value, which are not as close to each other
as observed for the EV parameter.  We believe that this is a
limitation of the maximum $N$ that we have considered in the
extrapolation, and a higher $N$ (which is difficult to simulate
presently) should lower the margin of difference in the extrapolated
value.  In the results to follow, we indicate this margin through an
error-bar in the data points.

The complete crossover of the two swelling ratios is shown in
\Figref{fig:aIHz} and is compared with that of \alfag.  It is observed
that $\alfaI \approx \alfag$ within error bounds for a large range of
$z$, i.e., the cross-over of static size measures is nearly identical.
This is a new result, calculated exactly for the first time for the
continuum Edwards model.  This is in contrast to the simple scaling
arguments of the thermal blob model \citep{WeiClo79}.  It is clear
that the ``spatial-crossover'' explanation of the blob model is
insufficient to show the observed difference in the crossover of the
swelling of hydrodynamic radius.  On the other hand, we see from
\Figref{fig:aIHz} that the crossover obtained from the long-time
diffusivity is significantly different and $\alfaH < \alfag$ for all
values of $z$.  Recognition of this fact is important since most of
the theoretical calculations are based on the static size measure
(\Dk) ignoring dynamic correlations ($D_{1}$).

\begin{figure}
  \centering
  \xlabel{$z$}
  \ylabel{$\alfaH; \alfaI$}
  \psfrag{aI}{\alfaI}
  \psfrag{aH}{\alfaH}
  \gpfig{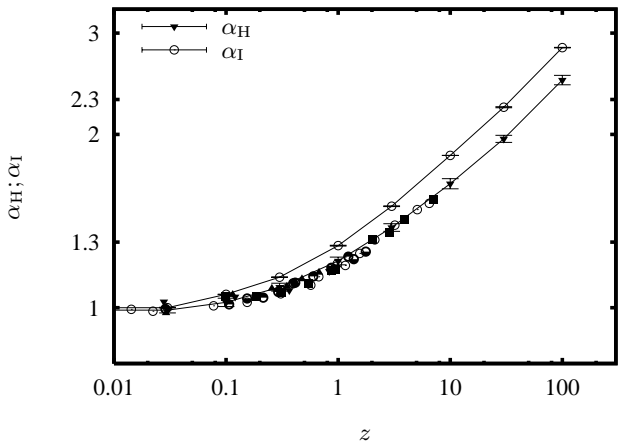}
  \caption{Crossover of hydrodynamic radius: Comparison of theoretical
    swelling of static and dynamic measures (line with points) with
    experimental data (points) collated in \Refcite{Tometal02}.}
  \label{fig:aIHecomp}
\end{figure}

\begin{figure}[tbph]
  \centering
  \xlabel{$\alfag$}
  \ylabel{$\alfaH; \alfaI$}
  \psfrag{aI}{\alfaI}
  \psfrag{aH}{\alfaH}
  
  \gpfig{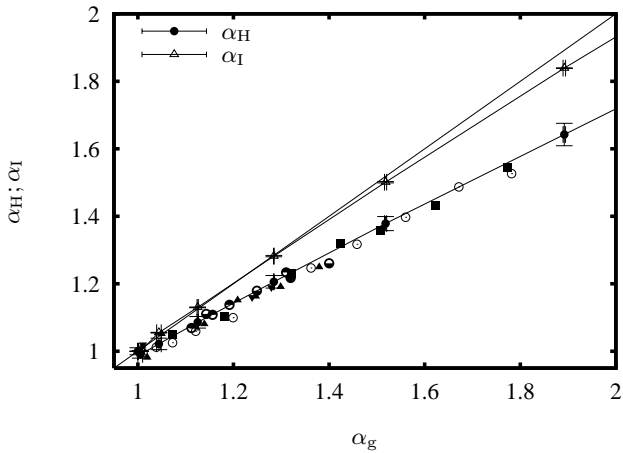}
  \caption{Parameter free comparison with measured values of \alfaH\
    against \alfag. Experiments are several data collated in
    \Refcite{Tometal02}.} 
  \label{fig:aHg}
\end{figure}

The theoretical crossover behaviour can be compared with experiments
in two ways.  In one method we can compare the $z$-crossover of the
swelling ratios, for which it is required to estimate the
phenomenological parameter $z$ for the experimental system, which has
a chemistry dependent prefactor.  In the alternative approach, we can
directly compare with two experimentally measured quantities, without
the need to estimate $z$.  We have chosen the most recent data (to our
knowledge) on the hydrodynamic radius, collated in
\Refcite{Tometal02}, in terms of a solvent quality they denote as
$\tilde{z}$.  We found that by taking $z = 0.85 \, \tilde{z}$, we can
shift the experimental data for the swelling of radius of gyration
\alfag\ on to our theoretical curve.  We then compare the crossover of
\alfaH\ using \emph{the same shift factor}.  \Figref{fig:aIHecomp}
shows this comparison leads to remarkable agreement.  It is also observed
that the crossover of \alfaI\ lies outside the region of experimental
accuracy.  This provides a substantial evidence that dynamic
correlations are important to describe the observed crossover behavior.  We
have also made a parameter-free comparison with the experimental data,
as shown in \Figref{fig:aHg}, for the measured values of \alfaH\ 
against that of \alfag, leading us to the same conclusion.

Based on the limited set of points shown in \Figref{fig:aIHz},
effective exponents for the static and dynamic properties can be
estimated at $z=10$, which is typically the largest experimental value
of $z$.  Taking $z \propto \sqrt{M}$, we find $\Rg \sim M^{0.59}$,
$\Ri \sim M^{0.58}$, and $\Rh \sim M^{0.57}$.  These results should be
taken only as indicative values of the exponent near $z = 10$, and are
not conclusive critical exponents.

To summarize,  Brownian  Dynamics simulations of the continuum
Edwards model  were used to study the crossover behavior
of the hydrodynamic radius  of a polymer chain. We found that the
commonly used static measure of diffusivity (obtained from the
inverse radius \Ri) has nearly identical crossover to that of
the gyration radius.  It is required to consider the long time
diffusivity (which has dynamic correlations) to explain the
experimentally observed slow crossover of dynamic properties.
Analytical calculations would be helpful to obtain further
insight into the results obtained here by simulations.

\begin{acknowledgements}
This work has been supported by a grant from the
Australian Research Council, under the Discovery-Projects program.
The computations were carried out in the facilities provided by the
Australian Partnership for Advanced Computation (APAC), and we are
grateful for their support.
\end{acknowledgements}


\end{document}